A Quantum Circuit to Construct All Maximal Cliques Using Grover's Search Algorithm

Chu Ryang Wie

Department of Electrical Engineering, University at Buffalo, State University of New York, Buffalo, NY
Email: wie@buffalo.edu

November 14, 2017

Abstract

A quantum circuit to construct all *maximal cliques* using *Grover's* search algorithm is presented. This oracle circuit takes as input an *n*-qubit state *|x>* and the adjacency matrix data *A* of an *n*-node network, and outputs the state $(-1)^{f(x)}|x>$ where *f(x)=1* if *|x>* is a maximal clique, and *f(x)=0* otherwise. The oracle workspace requires $2n^2$ qubits. Each oracle call makes about $10n^2$ calls to the *Toffoli* gates. The overall *Grover's* search algorithm takes as *input* a uniform superposition *n*-qubit state and the adjacency matrix data, and *outputs* a uniform superposition of all *maximal clique states* in time $O(\sqrt{n^2 2^n/M})$, where *M* is the total number of maximal cliques in the network.

# 1    Introduction

A quantum algorithm can achieve a significant speedup over a classical algorithm by taking advantage of quantum parallelism and quantum entanglement. The quantum parallelism can be achieved if a circuit on *n* qubits forms a linear superposition of the $2^n$ basis states and manipulates its $2^n$ amplitudes in parallel. Quantum entanglement enables a quantum algorithm to measure or to process further on a set of qubits the results computed on a different set of qubits. A clever use of quantum parallelism and entanglement enabled a significant speedup for certain types of computational problems: for example, an exponential speedup in the factoring algorithm [1] and phase-estimation algorithm [2] (part of the 'hidden subgroup' problems [3, 4, 5]) and a quadratic speedup in the Grover's quantum search algorithm (one of the 'amplitude amplification' algorithms) [6, 4, 5].

Recently, Lloyd et. al. presented a topological quantum algorithm to analyze big data [7]. They mapped vectors, simplices, simplicial complexes and collections of simplicial complexes to quantum states and designed a topological algorithm on the simplicial complexes built by the Grover's search algorithm to extract the persistent homology of data as a function of the grouping scale, the metric distance between data points. Topological features of big data or complex networks are robust to noise, to the way data are sampled, or to how they are represented. A data consisting of *n* data points possesses $2^n$ possible subsets. Likewise, a network consisting of *n* nodes possesses $2^n$ possible subsets of nodes. Hence, in a topological analysis, where each subset can represent a clique or a simplex, there exist an exponentially increasing number of subsets with an increasing number of nodes. In a quantum algorithm, however, each node can be represented by a qubit. For example, an *n*-qubit state $|x> = |x_1 x_2..x_j..x_n>$ can represent a subset where $x_j = 1$ indicates the node or data point *"j"*, and all the possible subsets can be processed in quantum parallel. By assuming that every pair of nodes in the subset is connected by an edge, $|x>$ represents a clique where a clique in a graph is an all-to-all connected set of vertices, or by letting every clique be filled in, $|x>$ represents a simplex. All possible cliques or simplices based on *n* nodes can be represented by a uniform superposition of *n*-qubit states. A quantum algorithm can provide a distinct speedup in the topological analysis of big data or complex networks and an exponential savings in the memory requirement [7].

Giusti et al analyzed the homology of a clique complex obtained from a weighted symmetric, neural correlations matrix *M* of a rat hippocampus [8]. From *M*, they obtained the so-called *order complex*, which is basically a sequence of symmetric, binary adjacency matrices *A's* (representing a sequence of undirected binary graphs) with a decreasing threshold value applied to the matrix entry of *M*. Each *A* is obtained by binarizing the weighted correlation matrix *M* by setting the matrix element to *1* if it is above a certain threshold value and *0* if below. The order complex may be obtained roughly as follows: starting with a threshold value just above the highest matrix element value in *M*, we have all elements of *A* equal to *0*. Next, lowering the threshold value to just below the highest matrix element value in *M*, the next *A* is obtained (likely with only one edge in the graph). Successively lowering the threshold value and binarizing *M*, a new *A* is obtained at each step. This sequence of binary adjacency matrices may be considered the *order complex* of ref.[8]. The undirected graph corresponding to an *A* obtained from *M* with a lower threshold value would have a higher edge density. From the clique complex obtained from each binary graph, they calculated the homology groups and the Betti numbers $\beta_m$ of order $m = 1, 2$ or $3$, which were then plotted as a function of the edge density of the graph. This yielded reliable and robust topological information about the rat's brain network. The Betti curves ($\beta_1$, $\beta_2$ and $\beta_3$ *vs.* the edge density) gave information about the geometry or randomness of the underlying hippocampal circuits [8]. By order complex, they took advantage of the fact that the ordering of matrix entries, irrespective of their actual values, carries information about the underlying matrix organization, and calculated the homology groups of the simplicial complex on the binarized matrix *A* as a function of the changing (lowering) threshold value for the matrix entry. This approach is similar to the approach of persistent homology where the changing variable is the metric distance between the pair of nodes. In the order complex, Betti numbers can be plotted as a function of the edge density $\rho$, showing a curve $\beta_m(\rho)$ for each dimension *m* (producing the Betti curve) [8]. Whereas, in the persistent homology [9], the Betti numbers of a various order can be plotted as a function of distance between nodes, resulting in a barcode-like plot [10]. The simplicial homology groups, $H_m$, can be calculated for each binarized adjacency matrix as a function of the edge density $\rho$. They called this a "clique topology" [8]. The Betti curve $\beta_m(\rho)$ can summarize the topological features of the weighted pairwise correlation matrix *M*. They suggest that the clique topology is especially useful in the biological settings for detecting structure in the presence of unknown nonlinearities [8].

We are interested in a quantum algorithm that can be used to analyze the topological features of human connectome data. The human connectome, which describes how various parts of human brain are connected, can be huge in size and complexity. In the human brain network the nodes may be defined by neurons, gray matter voxels or brain regions, and the links are defined by synapses, anatomical fiber bundles or dynamic coupling, representing measures of structural or functional connectivity [11, 12, 13]. The neural activity or connectivity data are often presented as a matrix whose entries $M_{ij}$ indicate the strength of correlation or anatomical connectivity between a pair of neurons, cell types or imaging voxels. A human brain network reported in literature [12] ranges from 70-node [14] to 140,000-node whole brain networks [15]. A complete brain network would have each node represent an individual neuron and each edge a synapse. Human brain has 100 billion neurons, each with about 7000 synapses [16]. Therefore, it is presently not practical or possible to image, or to analyze computationally, a complete human connectome. Only the most simple organisms, like worm *Caenorhabditis elegans* which has 302 neurons in total, have a complete connectome at the neuron-synapse level [17, 18, 19], which was recorded recently for a real time neuronal activity in three dimensions, enabling them to watch neurons firing in the brain, ventral cord, and tail [20]. Even insect brains have only a partial map (at neuronal level). A recent study of a fruit fly larva brain showed a complete connectome of a learning and memory centre involving roughly 1,600 of the 10,000 neurons contained in the larva's entire brain [21]. An adult fruit fly brain has roughly 100,000 neurons.

The study of human connectome will benefit from a topological analysis of the network in a quantum computer. A quantum oracle circuit to generate all maximal cliques from a binarized adjacency matrix *A* is useful for such quantum topological algorithms. We report in this paper a concrete quantum circuit that fulfills such a goal. This quantum circuit can serve as the oracle in the Grover's search algorithm (see *Appendix-B*). In the uniform superposition of $2^n$ states, corresponding to the $2^n$ subsets of *n* nodes, this oracle identifies all *maximal* clique states according to the binary adjacency matrix *A*, but it does not report cliques that are faces of a larger clique.

## 2     Constructing the Quantum Circuit

We wish to test if a candidate clique is *maximal* in the adjacency matrix data. Each term /x> in the uniform superposition of all states $\frac{1}{\sqrt{N}}\sum_{x=0}^{N-1}|x\rangle$ is considered a clique, where $N=2^n$ and *n* is the number of qubits or nodes. Writing /x> = /$x_1x_2..x_n$> in a binary form, a non-zero $x_j$ represents the node *"j"* where *j=1, 2, .., n*. Each /x> is tested, in quantum parallel, against the adjacency matrix data *A*.

In order to generate all maximal cliques in a basic classical algorithm [22], one starts with an undirected graph *G*, given as the symmetric binary matrix *A*. A clique is a complete subgraph of *G*, *i.e.*, a subgraph where any two vertices are adjacent. We first summarize the terminology relevant to cliques and graph theory according to ref.[23].

A simple undirected graph $G=(V,E)$ is considered for a finite set $V$ of *vertices* or *nodes* and a finite set $E$ of *unordered* pairs $(v,w)$ of distinct vertices, called *edges*. A pair of vertices $v$ and $w$ are *adjacent* if $(v,w) \in E$, and the nodes $v$ and $w$ are called *neighbors*. For a vertex $v \in V$, $N(v)$ is a set of all vertices adjacent to $v$: $N(v) = \{w \in V | (v,w) \in E, w \neq v\}$. For a subset of vertices $W \subseteq V$, $G(W)=(W,E(W))$ with $E(W) = \{(v,w) \in W \times W | (v,w) \in E\}$ is called a *subgraph induced by W*. The number of elements in $W$ is denoted by $|W|$. Given a subset $Q$ of vertices, $Q \subseteq V$, the induced subgraph $G(Q)$ is said to be *complete* if $(v,w) \in E$ for all $v,w \in Q$ with $v \neq w$, and $G(Q)$ is called a *complete subgraph*. A complete subgraph is a *clique*. If a clique is not a proper subgraph of another clique, then it is called a *maximal clique*.

A basic framework of classical algorithm to generate all maximal cliques is as follows: let $Q=\{p_1, p_2, ..., p_d\}$ be a complete subgraph (i.e., clique) at some stage. If $N(p_1) \cap N(p_2) \cap ... \cap N(p_d) = \emptyset$, then $Q$ is a maximal clique, else add to $Q$ one of the nodes that are a common neighbor to every node in $Q$ and repeat. The intersection operation $N(p_1) \cap N(p_2) \cap ... \cap N(p_d)$ produces all nodes, other than the nodes already in $Q$, that are neighbors to every node in $Q$. This basic approach may be directly applied to a quantum algorithm, in quantum parallel.

In order to build the quantum oracle circuit (for the Grover's search algorithm, *Appendix-B*), let $Q=\{p_1, p_2, ..., p_d\}$ be a *clique*. Then, $Q$ is a *maximal* clique if $(N(p_1) \cup \{p_1\}) \cap (N(p_2) \cup \{p_2\}) \cap ... \cap (N(p_d) \cup \{p_d\}) = Q$. (See *Appendix-A* for a Matlab code.) For each node $p_j$, $N(p_j) \cup \{p_j\}$ corresponds to the $j^{th}$ column vector $|C_j\rangle$ of $I+A$:

$$I + A = \sum_{j=1}^{n} |C_j\rangle \langle 0..0x_j0..0|$$

Here, $x_j=1$ and

$$|C_j\rangle = |c_{1j}c_{2j}\cdots c_{nj}\rangle = \begin{pmatrix} c_{1j} \\ c_{2j} \\ \vdots \\ c_{nj} \end{pmatrix}$$

where $c_{ij}$ for $i,j=1,2,...,n$ is the matrix element of $A+I$. The *maximal* clique candidate $Q$ is represented by a quantum state $|x\rangle=|x_1x_2...x_n\rangle$ where by definition, every pair of nodes is connected by an edge. The node "$j$" is encoded by $x_j=1$ in the state $|x\rangle=|x_1...x_j..x_n\rangle$ of the $n$-qubit quantum register. That is, the node "$j$" is *present* if $x_j=1$ and *absent* if $x_j=0$ in the state, and the state $|x\rangle$ is a term in the uniform superposition of all possible (maximal clique) states, $\frac{1}{\sqrt{N}}\sum_{x=0}^{N-1}|x\rangle$.

To test whether the state $|x\rangle=|x_1x_2..x_n\rangle$ is a maximal clique in the graph, we only need to take a bitwise multiplication of the matrix $A+I$ column vectors for every node in $|x\rangle = |x_1x_2..x_n\rangle$. This is equivalent to the *intersection* of the node sets where each set consists of a node in $|x\rangle$ and all of its neighbors. The *intersection* operation produces a set consisting of all common neighbor nodes. A general formula for set *intersection* operation between the subsets of nodes represented by the column vector $|C_j\rangle$ in the matrix $A+I$ is as follows: Given $|x\rangle=|x_1...x_j..x_n\rangle$,

$$\text{Intersect}(\{|c_{1j}c_{2j}..c_{nj}\rangle|\ \forall x_j=1, j=1,..,n\}) = |c_1c_2..c_n\rangle, \quad \text{where } c_i = \prod_{j=1}^{n} c_{ij}$$

The oracle circuit therefore consists of three steps. The first is to find the common neighbors (where the neighbors are in the column or row vector of the binary adjacency matrix $A$) for all nodes in the candidate clique $|x\rangle$. This is done by deselecting all column vectors for the nodes corresponding to $x_j=0$. We deselect the column vector by changing all of its entries to $1$ so that the intersection operation is not affected by this column vector. The second step is to perform the set *intersection* operation among all column vectors $|C_j\rangle = (I+A)|0..0x_j0..0\rangle$ for every node present in the candidate clique. Here, $I$ is an $n \times n$ *identity* matrix which is to include the node *"j"* itself along with its neighbors from the adjacency matrix $A$. The resulting vector from the *intersection* operation is equal to $|x\rangle$ if $|x\rangle$ is a maximal clique according to $A$. The *intersection* operation is achieved by a *controlled-not* operation $C^n(X)$ for $n$ nodes. The third step compares the resulting vector with the candidate state $|x\rangle$.

**The oracle quantum circuit**: The common neighbors of all nodes in $|x\rangle = |x_1,..,x_j,..,x_n\rangle$ are found by a *controlled-not* $C^n(X)$ operation where each qubit in the ancilla state acts as a control (for *bitwise-AND*) and the candidate clique state $|x\rangle$ acts as the target. This completes the second and third steps above. Here, the column vectors $|C_j\rangle$ with $x_j=0$ should have no effect. Hence, the following steps are taken before the *bitwise-AND* or the set *intersection* operation: (1) the bit values of column vector state $|C_j\rangle$ with $x_j=1$ are copied onto the corresponding bits of the ancilla state; and (2) for $|C_j\rangle$ with $x_j=0$, every bit is set to $1$ on the ancilla state, independent of the bit value in $|C_j\rangle$. Therefore, the ancilla bits are determined according to the following formula:

$$x_j c_{ij} \oplus \bar{x}_j c_{ij} \oplus \bar{x}_j \bar{c}_{ij} = \begin{cases} c_{ij} & \text{if } x_j = 1 \\ 1 & \text{if } x_j = 0 \end{cases}$$

where $\oplus$ is the Boolean *XOR* operator. This formula was implemented using three variants of the *Toffoli* gate with $x_j$ and $c_{ij}$ acting as controls and an ancilla bit as the target. This is shown in the Figure 1 for a network with three nodes.

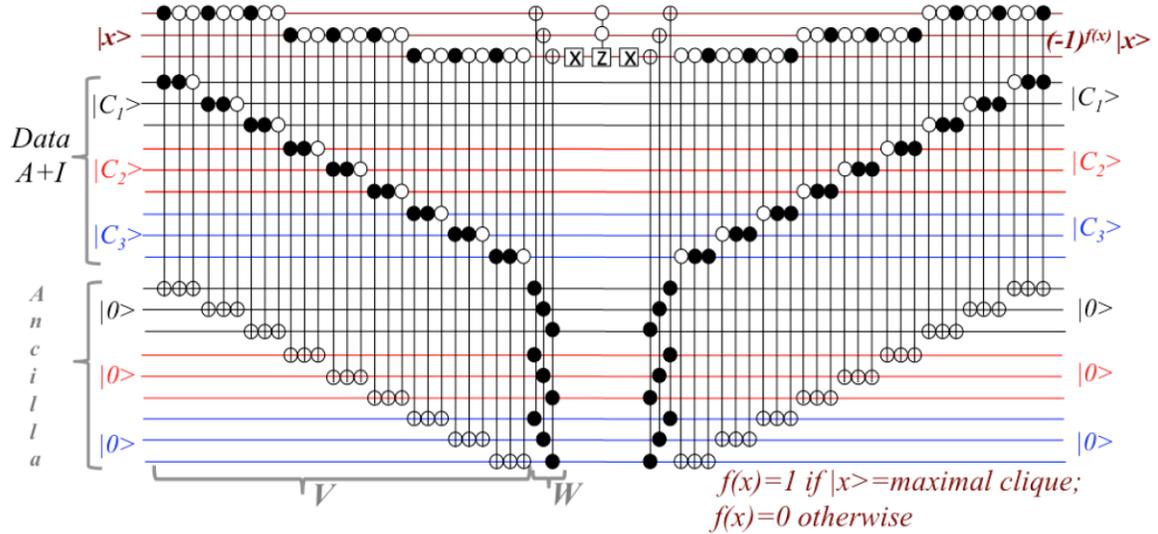

$f(x)=1$ if $|x\rangle$=maximal clique;
$f(x)=0$ otherwise

Figure 1   The oracle circuit to construct all maximal cliques in a network with $n=3$ nodes. The first $n$ qubits at the top represent the potential maximal clique where the state $|x>$ is a member of the uniform superposition state of all possible states.  The $2n^2$ qubits below are the oracle workspace consisting of $n^2$ data qubits and $n^2$ ancilla qubits. (See *Appendix-B* for review of the Grover's search algorithm which uses this oracle.)

## 3   Discussion

In the circuit of Fig.1, the unitary gate *V*, which is a group of $3n^2$ Toffoli gates where $n$ is the total number of nodes in the network ($n=3$ in Fig.1), copies the bit values of the column vector $|C_j>$ onto the corresponding ancilla qubits if the node "*j*" is present (*i.e.*, the bit $x_j=1$) in the candidate clique $|x>$; but if $x_j=0$, every bit in the target ancilla is set to *1*.  The unitary gate *W* is a group of $n$ $C^n(X)$ gates.  In each $C^n(X)$ gate, $n$ ancilla bits are the control and a bit in the *maximal clique* candidate state $|x>$ is the target.  They perform the *intersection* operation for the sets of adjacent nodes, the columns of matrix *A+I*, and compare the result of intersection with the clique state $|x>$, turning it to $|0>$ if the result equals $|x>$.

For example, consider a graph of three nodes below with its adjacency matrix *A*. Let us test if $|110>$, which is {*1, 2*}, and $|111>$, or {*1, 2, 3*}, form a maximal clique or not. From the graph we know that $|110>$ is a maximal clique, and $|111>$ is not.

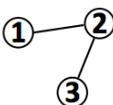

$$A = \begin{pmatrix} 0 & 1 & 0 \\ 1 & 0 & 1 \\ 0 & 1 & 0 \end{pmatrix}, \quad I + A = \begin{pmatrix} 1 & 1 & 0 \\ 1 & 1 & 1 \\ 0 & 1 & 1 \end{pmatrix}$$

For the clique candidate $|x>=|110>$, we form the set of common neighbors by the set *intersection* operation of the columns *1* and *2* of the matrix *I+A* (note: column numbers run *1* through *3*).  The resulting common-neighbor set is {*1, 2*}, *i.e.*, the state vector $|110>$.  That is, using $|C_j>$ to represent the $j^{th}$ column vector of *I+A*,

$Intersect(|C_1>, |C_2>) = Intersect(|110>, |111>) = |110> = common\ neighbor\ \{1,2\}$

A $C^3(X)$ operation from this state $|110>$ (which is the result of a *bitwise-AND* or *Control*$^3$ of the three ancilla states) to the clique candidate state $|x>=|110>$ (the target state of $C^3(X)$) yields $|000>$.  The sign of the state $|000>$ is flipped to $-|000>$ by the unitary gate $I-2|0><0|$, which follows the *W* gate, in the top register in Fig.1.

For the candidate state $|x>=|111>$, the *intersection* of columns 1, 2 and 3 of the matrix *I+A* yields a common neighbor set {*2*}, *i.e.* $|010>$:

$Intersect(|C_1>, |C_2>, |C_3>) = Intersect(|110>, |111>, |011>) = |010>$

Hence, $C^3(X)$ from this result $|010>$ to the candidate state $|x>=|111>$ yields $|101>$, and the unitary operator $I-2|0><0|$ does not flip the sign of the candidate state.

The oracle workspace requires $2n^2$ qubits consisting of $n^2$ data qubits of the *A+I* data matrix and $n^2$ ancilla qubits.  Each oracle call requires execution of $6n^2$ Toffoli gates (*V* and $V^\dagger$), $2n$ $C^n(X)$ gates for the set intersection and state comparison operations (*W* and $W^\dagger$), and a single call to the operator *I-2|0><0|*, which flips the sign of state *|0>* relative to all other *|x>* states (where *x≠0*), and is implemented by *2n X*-gates and one $C^{n-1}(Z)$ gate. Some *Toffoli* gates in the unitary *V* require additional single-qubit *X* gates on the control bits. It is worth noting that a $C^n(U)$ gate can be implemented by *2(n-1) Toffoli* gates and a single two-qubit *controlled-U* gate and requires additional *n* ancilla qubits for workspace [4], where *U* is a single qubit unitary. Therefore, this maximal-clique oracle circuit has a computational complexity polynomial in *n*, whereas the complexity of overall Grover search algorithm is $O(\sqrt{n^2 2^n/M})$ to build all maximal cliques (see *Appendix-B*). In a classical algorithm, the worst-case time complexity is $O(3^{n/3})$ to build all maximal cliques for an *n*-vertex graph in a modified version of the Bron-Kerbosch algorithm by Tomita et al. [23]. This indicates that the quantum Grover search algorithm provides little speedup over the classical algorithm. However, the output state *|χ>*, the uniform superposition of all maximal cliques, obtained as the output of the quantum search algorithm, can serve as input to a quantum topological analysis algorithm similar to that given in ref.[7].

## 4      Summary

A quantum oracle circuit to construct all maximal cliques was presented. This circuit can function as the oracle for Grover's search algorithm whose output is a uniform superposition of all maximal clique states according to the adjacency matrix data of the network, obtain in time $O(\sqrt{n^2 2^n/M})$ where *n* is the number of nodes and *M* is the total number of maximal cliques. This uniform superposition state of maximal cliques can be used as input to a quantum topological analysis algorithm.

# Appendix

## A  Matlab code for testing if $x_1x_2..x_n$ is a maximal clique in the adjacency matrix *A*

Matlab code (classical)
For every $x(decimal)=x_1x_2..x_n(binary)$ and for a given adjacency matrix *A*, this Matlab code checks every column vector of *A*, and builds a common-neighbor *node set* using the *set intersection* operation for the node sets from column *"j"* of *A* for every $x_j=1$ in the $x=x_1..x_j..x_n$. A row vector may be used instead of column vector for the symmetric *A*.

Matlab code:
```
for x=2^n-1:-1:1
    xb=dec2bin(x,n)-'0';   %x_b = x_1x_2 .. x_n
    nodesetx=find(xb); %find x_j, or the position "j" with non-zero elements.
    nodesetA = 1:n;  %start with every node, from 1 to n.
    for j=nodesetx   %for each node "j" in the candidate clique x_b
        nodesetA=intersect(nodesetA, union(j,find(A(j,:))));
        if ~isempty(setdiff(nodesetx,nodesetA));
            break; %we only want a "maximal" clique nodesetx
        end
    end
    if isequal(nodesetx, nodesetA)
        newMC=zeros(1,n);
        newMC(nodesetx)=1; %new maximal clique
        MC=[MC newMC.']; %add to the collection of maximal cliques.
    end
end
```

# B     Review of Grover Search Algorithm, Grover Operator, and Counting Algorithm (Phase Estimation)

## B.1     Grover's Search Algorithm and Grover Operator G

The uniform superposition of all states $|\psi\rangle$ is a linear combination of the marked states (i.e., every maximal clique state $|x\rangle$ with $f(x)=1$), represented collectively by a normalized state $|\chi\rangle$, and the unmarked states (i.e., all other states $|x\rangle$ with $f(x)=0$), represented by a normalized state $|\xi\rangle$.

$$|\psi\rangle = \frac{1}{\sqrt{N}} \sum_{x=0}^{N-1} |x\rangle = \sqrt{\frac{M}{N}} \frac{1}{\sqrt{M}} \sum_{x,f(x)=1} |x\rangle + \sqrt{\frac{N-M}{N}} \frac{1}{\sqrt{N-M}} \sum_{x,f(x)=0} |x\rangle$$

$$= \sin\frac{\theta}{2}|\chi\rangle + \cos\frac{\theta}{2}|\xi\rangle$$

where $M$ is the total number of maximal cliques (the marked states), and

$$\sin\frac{\theta}{2} \equiv \sqrt{\frac{M}{N}}, \quad |\chi\rangle \equiv \frac{1}{\sqrt{M}} \sum_{x,f(x)=1} |x\rangle, \quad \text{and } |\xi\rangle \equiv \frac{1}{\sqrt{N-M}} \sum_{x,f(x)=0} |x\rangle$$

The oracle $O$ changes each state $|x\rangle$ to $(-1)^{f(x)}|x\rangle$. Therefore, $O|\chi\rangle = -|\chi\rangle$ and $O|\xi\rangle = +|\xi\rangle$, and

$$O|\psi\rangle = \frac{1}{\sqrt{N}} \sum_{x=0}^{N-1} (-1)^{f(x)} |x\rangle = -\sin\frac{\theta}{2}|\chi\rangle + \cos\frac{\theta}{2}|\xi\rangle$$

The oracle $O$ is followed by the 'inversion about the mean' operator $U$ defined as

$$U = 2|\psi\rangle\langle\psi| - I = H^{\otimes n}(2|0\rangle\langle 0| - I)H^{\otimes n}$$

The pair of operators, $O$ and $U$, constitutes the Grover operator $G$.

$$G \equiv UO = [H^{\otimes n}(2|0\rangle\langle 0| - I)H^{\otimes n}]O$$

Starting with the uniform superposition state $|\psi\rangle$, successively applying $G$ for a total of $R$ times we obtain,

$$G^R|\psi\rangle = \sin\frac{(2R+1)\theta}{2}|\chi\rangle + \cos\frac{(2R+1)\theta}{2}|\xi\rangle$$

In order to have only the maximal cliques (the marked states) as the output, we need $R$ to be

$$\frac{(2R+1)\theta}{2} \approx \frac{\pi}{2} \rightarrow R \approx \frac{\pi}{4}\sqrt{\frac{N}{M}} \quad \text{for } N \gg M$$

This leads to $G^R|\psi\rangle \approx \sin\frac{\pi}{2}|\chi\rangle + \cos\frac{\pi}{2}|\xi\rangle = |\chi\rangle$ as the output. The final output $|\chi\rangle$ is a uniform superposition state of all maximal cliques in the network. This is illustrated in the circuit of Figure B-1.

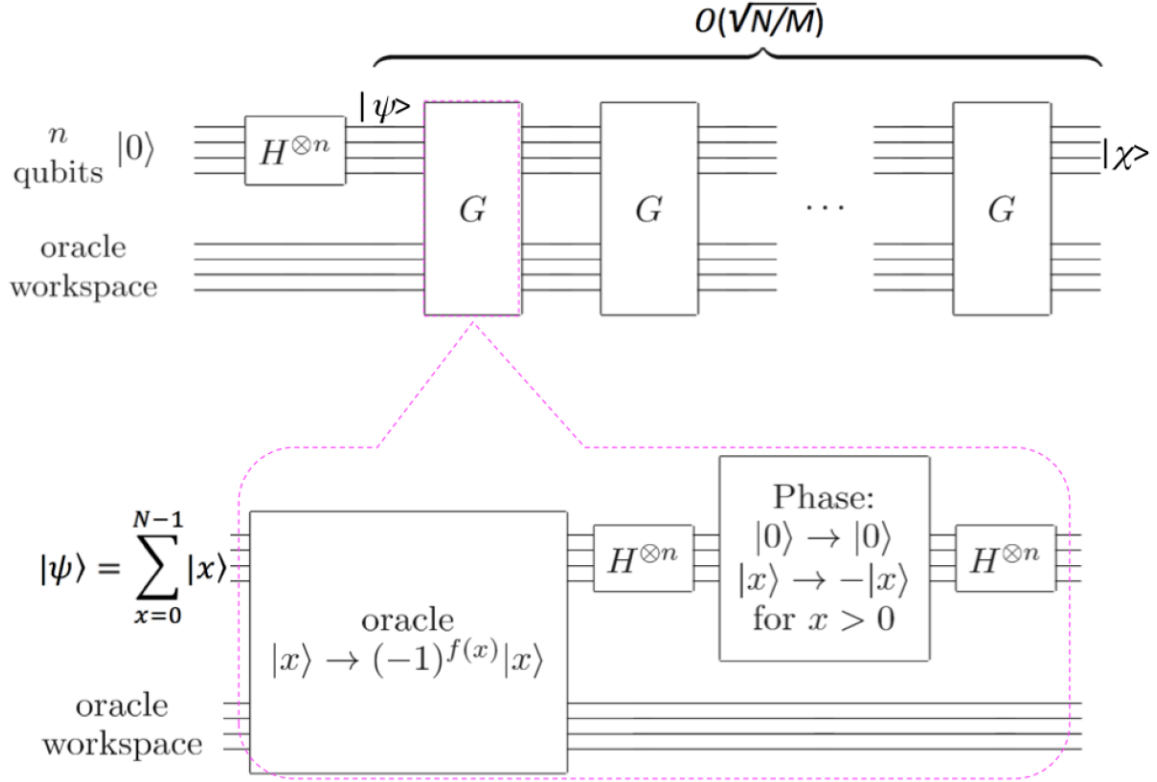

Figure B-1  Grover's search algorithm.  In this paper, a quantum circuit is given for the oracle to identify the maximal cliques: *f(x)=1* if *|x>* is a maximal clique; and *f(x)=0*, otherwise.

The Grover's algorithm requires knowledge of the total number of maximal cliques *M* in the network, so that the value *R*, the total number of Grover iterations, is known.  Only then can we construct the state $|\chi\rangle$, representing the set of all maximal cliques, by Grover's algorithm.

### B.2    Quantum Counting (Phase Estimation Algorithm)

Quantum counting algorithm is used to estimate the total number *M* of maximal cliques (the marked states) in the network.  We start by rewriting the uniform superposition state *|ψ>* in terms of the eigenstates $|\phi_\pm\rangle$ of the Grover operator G.

$$|\psi\rangle = \frac{e^{i\theta/2}|\phi_+\rangle + e^{-i\theta/2}|\phi_-\rangle}{\sqrt{2}}$$

where,

$$|\phi_\pm\rangle \equiv \frac{|\xi\rangle \mp i\,|\chi\rangle}{\sqrt{2}}, \quad and \quad G|\phi_\pm\rangle = e^{\pm i\theta}|\phi_\pm\rangle$$

The phase estimation algorithm is then applied with the uniform superposition state *|ψ>*, which includes the maximal cliques, as the target state and a new set of *t* qubits

(register-1) as the control qubits.  The quantum counting algorithm is depicted in Figure B-2.

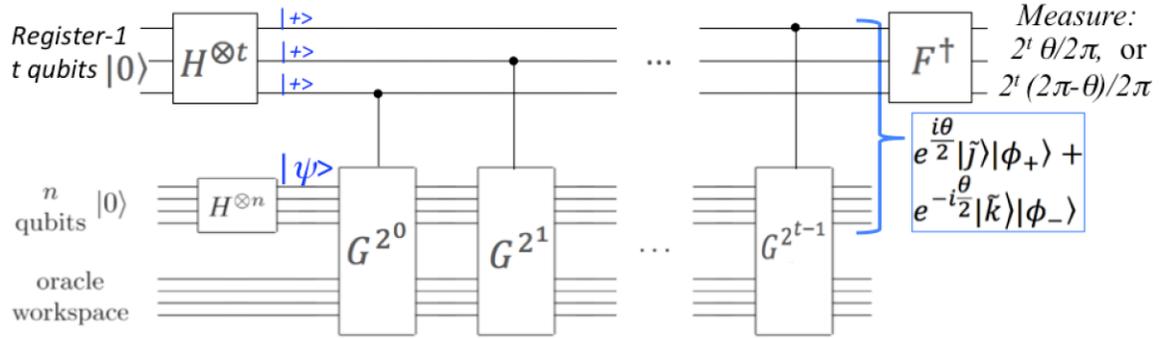

Figure B-2  Quantum counting algorithm to estimate $\theta$ or $2\pi$-$\theta$, in order to find the total number $M$ of all maximal cliques.

The quantum state just before the inverse quantum Fourier transform $F^\dagger$ is a superposition of the two Grover eigenstates $|\phi_+\rangle$ and $|\phi_-\rangle$ entangled with the states $|\tilde{j}\rangle$, the Fourier transform of $|j\rangle=|j_1j_2..j_t\rangle$, and $|\tilde{k}\rangle$, the Fourier transform of $|k\rangle=|k_1k_2..k_t\rangle$, respectively, where $j$ and $k$ are defined as
$$\theta \equiv 2\pi \frac{j}{2^t}, \quad 2\pi - \theta \equiv 2\pi \frac{k}{2^t}, \quad and \quad 0 \leq j, k < 2^t$$
This entangled state is
$$e^{i\theta/2}|\tilde{j}\rangle|\phi_+\rangle + e^{-i\theta/2}|\tilde{k}\rangle|\phi_-\rangle = e^{i\theta/2}|\widetilde{2^t\theta/2\pi}\rangle|\phi_+\rangle + e^{-i\theta/2}|\widetilde{2^t(2\pi-\theta)/2\pi}\rangle|\phi_-\rangle$$
After applying the inverse QFT $F^\dagger$, we measure all $t$ bits of register-1 to find either $j$ or $k$ to a $t$-bit accuracy, thereby giving an approximate value of $\theta$ or $2\pi$-$\theta$. They both yield the same $M$ value:  $M = 2^n \sin^2((2\pi$-$\theta)/2) = 2^n \sin^2(\theta/2)$.